\documentclass[aps,superscriptaddress,twocolumn,floatfix,citeautoscript,longbibliography,hyperlinks]{revtex4-2}

\usepackage{graphicx}
\usepackage{dcolumn}
\usepackage[dvipsnames]{xcolor}
\usepackage[colorlinks,citecolor=blue,urlcolor=blue]{hyperref}
\usepackage{amssymb,amsmath,bm,bbm,mleftright,mathtools}
\DeclareMathAlphabet\mathbfcal{OMS}{cmsy}{b}{n}

\renewcommand{\vec}{\mathbf}
\makeatletter
\newsavebox{\@brx}
\newcommand{\llangle}[1][]{\savebox{\@brx}{\(\m@th{#1\langle}\)}%
  \mathopen{\copy\@brx\kern-0.5\wd\@brx\usebox{\@brx}}}
\newcommand{\rrangle}[1][]{\savebox{\@brx}{\(\m@th{#1\rangle}\)}%
  \mathclose{\copy\@brx\kern-0.5\wd\@brx\usebox{\@brx}}}
\makeatother

\begin{document}

\title{Entanglement from particle number fluctuations in a second-order topological insulator}

\author{ Mahla Moridi Farimani}
\affiliation{Computational Physics Laboratory, Physics Unit, Faculty of Engineering and
Natural Sciences, Tampere University, FI-33014 Tampere, Finland}

\author{ Kim P\"oyh\"onen}
\affiliation{Computational Physics Laboratory, Physics Unit, Faculty of Engineering and
Natural Sciences, Tampere University, FI-33014 Tampere, Finland}

\begin{abstract}
Higher-order topological insulators are materials with a bulk energy gap featuring gapless modes at $(d-n)$-dimensional edges, where $n > 1$. Despite this modified bulk-boundary correspondence, their nontrivial topology can nevertheless be observed through the bipartite entanglement spectrum. Inspired by works suggesting fluctuations in conserved quantities could be used to study topology in one-dimensional systems, which likewise feature zero-dimensional edge modes, we extend this treatment to a two-dimensional second-order topological insulator where corner states play an analogous role. We show that the standard bipartite entanglement entropy and particle number fluctuation lack striking signals of topological phase transitions, both in the static case and in the time evolution after a quench, obscured by an area-law term from the subsystem edges. By introducing a quadripartite construction of the subsystem, we show that it is possible to isolate the topological contributions, producing sharp transition peaks in both quantities, observable by studying subsystems much smaller than the full system. As particle number fluctuations are accessible in existing experiments, these results establish a concrete and scalable path to experimental detection of the entanglement in higher-order topological insulators. \end{abstract}
\maketitle

\section{Introduction}

Entanglement has been one of the central concepts of quantum mechanics since the early days of the field \cite{EPR1935}. Due to its fundamental role it acts as a unifying concept drawing parallels between otherwise apparently disparate fields of physics, from quantum information to black holes \cite{Horodecki2009,CalabreseCardy2004,hayden2007,Tatsuma2018holography,Abanin2019rmp}.
In the theory of topological materials, entanglement has also played a central role: concepts such as entanglement entropy and the general features of the entanglement spectrum have been used to analyze both different topological phases and the various many-body phenomena that can occur related to them \cite{TEE_2006,wen2006,haldane2008,pollmann2010,jiang2012identifying,laflorencie2016quantum}. In particular, similarly to how the topological phase of a finite-size system is reflected in its energy spectrum through the bulk-boundary correspondence, it is also reflected in the bipartite entanglement spectrum of a subsystem, which will tend to have an entanglement gap closing where the system energy spectrum would have a gap closing\cite{Fidkowski2010}.
\begin{figure}[h!]
    \centering
    \includegraphics[width=1\linewidth]{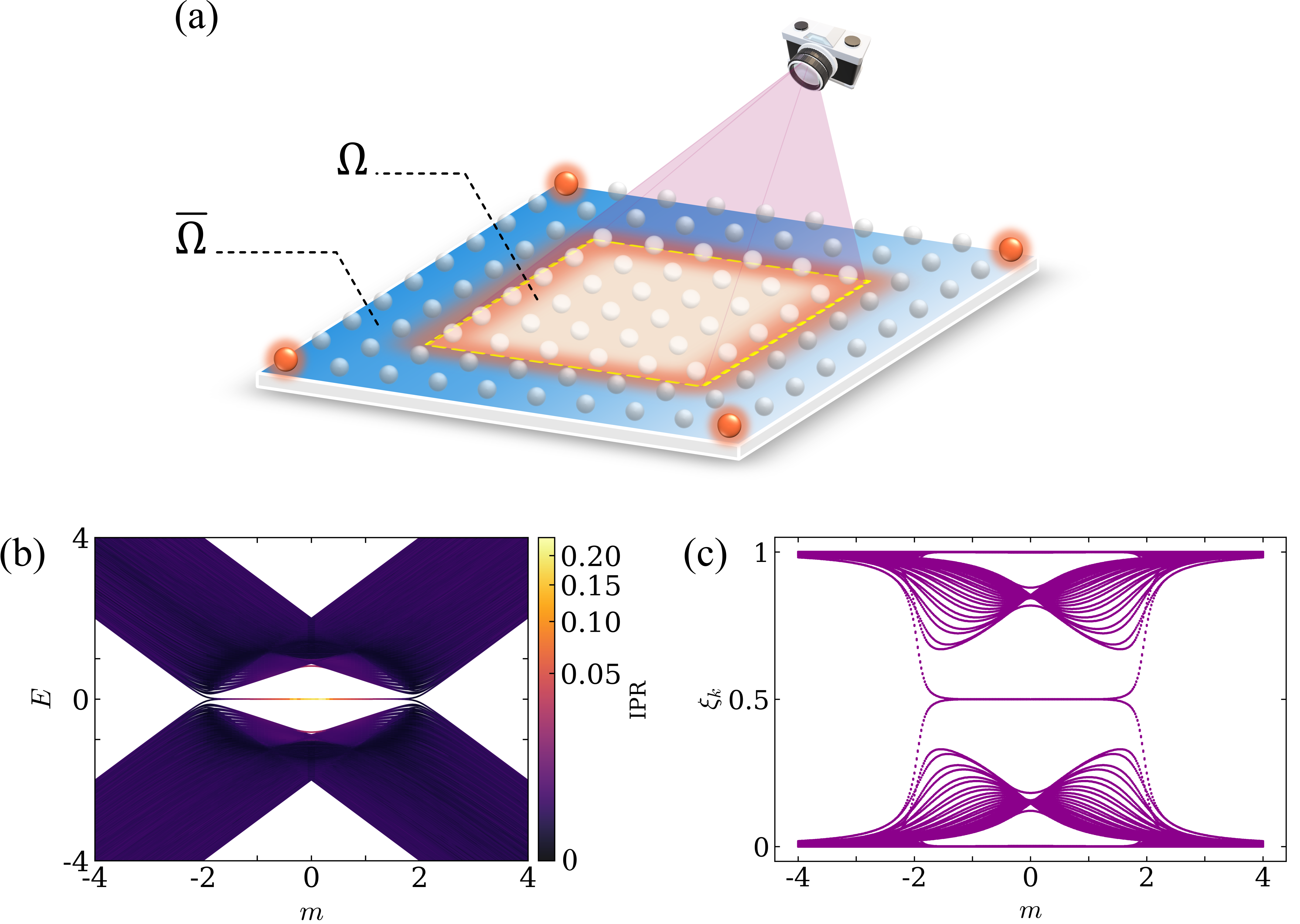}
    \caption{(a) Schematic figure; entanglement properties are obtained by choosing a subsystem away from the boundaries of the system and performing measurements on the subsystem. This corresponds to making use of the reduced density matrix.
    (b) Energy spectrum of the model used here as a function of the parameter $m$ for a system size $L=30$. At $m = \pm 2$ the gap closes signifying a transition to a second-order topological insulator when $|m| < 2$. The eigenvalues have been colored by the inverse participation ratio $\text{IPR} = \sum_{\vec r_j} |\psi({\vec r}_j)|^4$ to highlight the fact that the zero-energy modes are localized to the corners of the system. (c) Spectrum of the correlation matrix for a subsystem of the model as a function of $m$. The setup is similar to the schematic in Fig. (a), with system size $L = 30$ and subsystem size $L_{\Omega} = 15$, and as with the spectrum, a phase transition can be clearly observed at $m = \pm 2$.}
    \label{fig:picture1}
    \centering
\end{figure}

In the past decade, the theory of symmetry-protected topological phases has been extended to so-called higher-order topological insulators (HOTIs), where the usual bulk-boundary correspondence is slightly altered. Where in a usual $d$-dimensional topological phase, the bulk of the system is gapped while the $(d-1)$-dimensional surfaces feature gapless modes, an $n$th order topological insulator instead sees gapless behaviour on $(d-n)$-dimensional surfaces of a system; for example, a cubic second-order topological insulator will have gapless modes on its hinges. Despite this altered bulk-boundary correspondence, it has been found that the topology of the higher-order phase can likewise be mirrored in the bipartite entanglement spectrum  \cite{wang_2018,zhang_2026}. However, the spectrum in itself, while an important theoretical tool, is not straightforward to study experimentally. The bipartite entanglement entropy, that can be calculated directly from the spectrum and has frequently been used as a tool to assess topological entanglement, is likewise not always an experimentally accessible quantity, often requiring techniques such as tomography that scale poorly with system size, or else system-specific indirect methods \cite{Cramer2010,Greiner2015measuring,Hafezi2016,Hauke2016,Lin2024}. Fluctuations of conserved quantities, most conveniently particle number, have been proposed as a more experimentally viable approach to studying bipartite entanglement \cite{Song_2012,Calabrese_2012,rachel_2012}, and recent works on one-dimensional systems have suggested the use of such as a proxy of entanglement entropy when probing topology, both statically and through the time evolution after a quench \cite{poyhonen_2022,glodzik_2025,poyhonen_2021}. Because the corner modes of second-order topological insulators in 2D are similar to the edge modes of 1D topological insulators, it is reasonable to wonder whether a similar approach would work in this type of system. 

In this work, we study the von Neumann entanglement entropy and the particle number fluctuations in subsystems of two-dimensional second-order topological insulators, and in particular examine how the latter could be used to obtain information about topological phases in the system. We first examine the bipartite quantities, which have been found to work well to model topology in 1D systems, before moving on to consider modified quantities based on linear combinations of bipartite quantities as well as to the dynamical properties of fluctuations after a quench across a phase transition. We find that while both the static and dynamic bipartite quantities do appear to show features that correlate with the topological phase for this particular model, they are not reliable as an indicator of the topological phase; however, a suitable combination of bipartite quantities, essentially further subdividing a selected subsystem into a quadripartite construction, can be used to remove contributions from trivial boundary term and provide a fluctuation-based indicator of topology.

\section{Model and methods}

We choose as our example model the second-order topological insulator described by the Hamiltonian 
\begin{equation}
\begin{split}
    H = \sum_{\substack{j \\ a = x,y}}&c^\dagger_{\vec r_j}\left(t \tau_z + i\lambda\tau_x\sigma_a + \Delta \cos(2\theta_a) \tau_y\right)c_{\vec r_j + \vec{\hat e}_a} + \mathrm{H.c.}\\
    &+ m\sum_{j} c^\dagger_{\vec r_j}\tau_zc_{\vec r_j}\label{eq:hamiltonian_c4}
\end{split}
\end{equation}
where $\tau$ and $\sigma$ act in orbital and spin space respectively, and the angle $\theta_a$ represents the angle the unit vector $\vec{\hat e}_a$ makes with the $x$ axis. The parameters $t$ and $m$ represent, respectively, nearest-neighbour hopping and an onsite mass term, while $\lambda$ is an antisymmetric hopping term coupling spin and orbital degrees of freedom. This model was previously studied in a similar context by Wang \textit{et al.} in Ref.~\cite{wang_2018}, and due to its simple structure makes for a good case study. As noted in \cite{wang_2018}, when $\Delta = 0$ the model represents a (first-order) topological insulator, which is topologically nontrivial when $|m| < 2$. Letting $|\Delta| > 0$ breaks time reversal and opens an edge gap, but due to the directional dependence of the term, the sign of this gap , given the sites at $\lbrace\vec r_j\rbrace$ are placed on a square lattice, $C_4T$ symmetry is preserved and protects a second-order insulating phase. Hereinafter, we will generally use the parameter values $t = \lambda = 0.5$ and $\Delta = 0.25$, varying only the parameter $m$ in order to shift the topological phase of the system. The term gapping the edges does not shift the phase boundaries of the model, and as such the nontrivial phase can be found when $-2 < m < 2$, featuring four zero-energy modes at each of the corners of a system with open boundary conditions (OBC), as seen in Fig.~\ref{fig:picture1}(b). 

In order to examine the entanglement properties of this system, let us consider a bipartition into a square sized subsystem $\Omega$ and its complement $\bar \Omega$. We will assume the full system $\Omega \cup \bar \Omega$ has open boundary conditions to closer reflect experimental reality, though as seen in Appendix~\ref{app_C}, as long as the subsystem is well inside the bulk of the system, the differences are not significant. In general, information about the subsystem $\Omega$ can be obtained from the reduced density matrix, $\rho_\Omega = \mathrm{Tr}_{\bar \Omega}\rho$. In this case, however, as the Hamiltonian \eqref{eq:hamiltonian_c4} is noninteracting, the single-particle entanglement spectrum can be obtained from the correlation matrix $C_{mn} = \langle c^\dagger_m c_n \rangle$ with the operators limited to sites in $\Omega$ \cite{peschel_2003,peschel_2009}. The eigenvalues $\xi_i$ of this matrix lie in the interval $[0,1]$, with $\xi = \frac{1}{2}$ corresponding to the center of the entanglement gap; they can be viewed as occupation numbers of an entanglement Hamiltonian, with the ground state formed of states corresponding to $\xi > 0.5$, and are as such sometimes called entanglement occupation. The topology of the system is clearly visible in the spectrum of the correlation matrix, as seen in Fig.~\ref{fig:picture1}(c). From the same data, we can directly obtain relevant entanglement quantities such as the (bipartite) von Neumann entanglement entropy
\begin{equation}
    S_{\mathrm{vN}} = -\sum_{i}\left[\xi_i \ln \xi_i + (1-\xi_i)\ln(1-\xi_i)\right]
\end{equation}
However, despite the ease of calculation, like the entanglement spectrum, the entanglement entropy is also difficult to obtain experimentally, often requiring procedures such as quantum tomography that scale poorly with system size \cite{laflorencie2016quantum}. A more accessible quantity may be fluctuations in conserved quantities, which have previously been discussed as a substitute for entanglement entropy in one-dimensional systems \cite{Calabrese_2012,Song_2012, rachel_2012, poyhonen_2022, moghaddam_2023}. In particular, the Hamiltonian of this model conserves the total particle number $N$, whose fluctuations in terms of variance can be immediately obtained from the entanglement occupation through the relation
\begin{equation}
    \mathrm{Var} N = \sum_i (\xi_i - \xi_i^2).
\end{equation}
This quantity is typically more experimentally accessible than the entanglement entropy. As it previously has been found useful in considering the topology of one-dimensional topological phases, and the 0D corner modes of the HOTI model we have here bears some resemblance to those states, this hence proves a natural starting point for analyzing the topological phase of this model. While the subsystem particle number fluctuations cannot be expected to give the exact value of the entanglement entropy, they will generally scale similarly for a wide range of systems \cite{poyhonen_2022}, which is often enough to extract information about the underlying topological phase.

\begin{figure}
    \centering
    \includegraphics[width=1\linewidth]{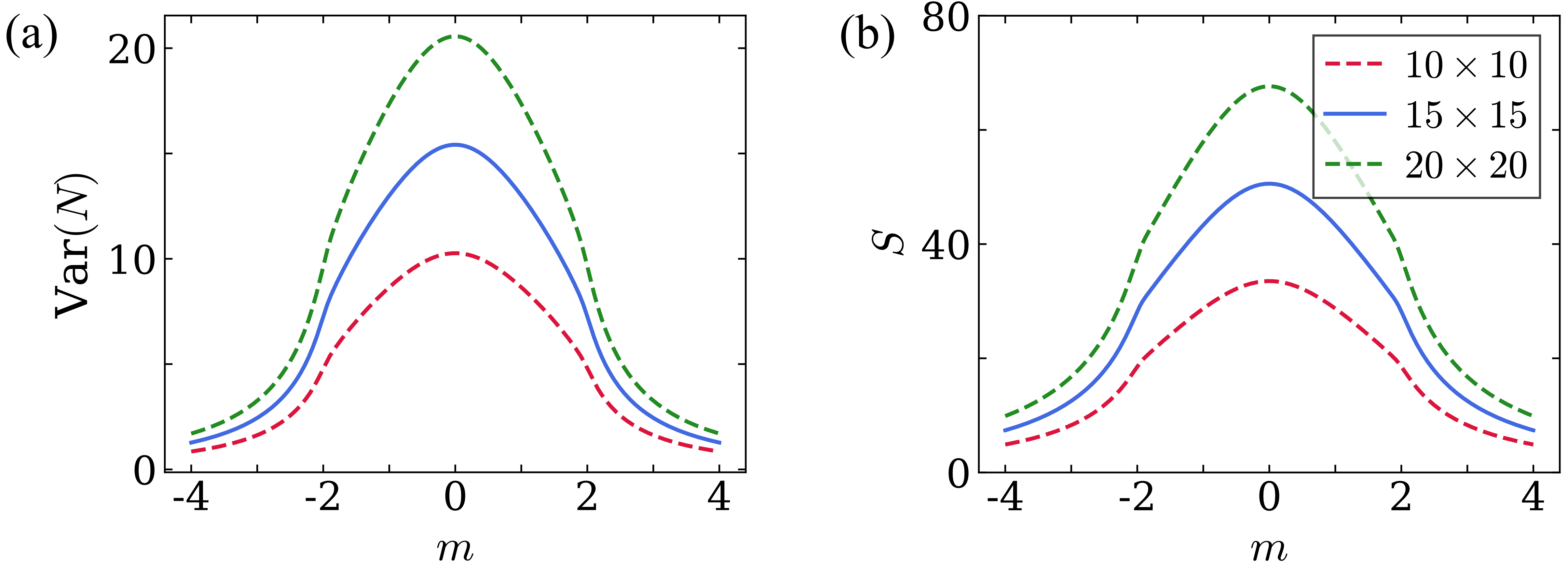}
    \caption{(a) Particle number fluctuation of system size $L = 30$ and three different subsystem sizes $(L_{\Omega}=10, 15, 20)$ as a function of $m$. (b) The von Neumann entanglement entropy for the same subsystems.}
    \label{fig:bipart_allsize}
\end{figure}

In Fig.~\ref{fig:bipart_allsize}, we show $S_{\mathrm{vN}}$ and $\mathrm{Var} N$ as a function of the mass parameter $m$, for a fixed subsystem $\Omega$ centered in a squared lattice for three different sizes (well represented by the schematic in Fig.~\ref{fig:picture1}). Both quantities are smooth, single-peaked functions of $m$, symmetric about $m=0$, and decay monotonically toward the trivial phase on either side. Notably, neither curve exhibits a kink, discontinuity, or local features at the transition points $m = \pm 4t$, despite the subsystem being well within the HOTI phase over an extended range of $m$. We do, however, observe that $S_{\mathrm{vN}}$ and $\mathrm{Var}$ track each other closely across the full range of $m$.

The absence of a sharp signal in $S_{\mathrm{vN}}$ and $\mathrm{Var}N$ is expected, as both quantities sum over the entire spectrum of the reduced correlation matrix of $\Omega$, which contains not only the non-trivial topological parts but also trivial contributions from the edges. The corner mode contributions, however, are quickly obscured by the trivial edge background as system size increases, and conversely behave poorly at small system sizes where the corner modes hybridize. Thus the topological information that is present in the bipartite entanglement occupation spectrum is mostly muddled by other states. It is therefore worth considering some other functions of the entanglement spectrum that could preserve the topologically important features but remove the contributions of the non-topological states, such as in Refs.~\cite{TEE_2006,Levin_2006}. This is precisely the motivation behind the quadripartite subtraction scheme of Ref.~\cite{wang_2018}, which is designed to cancel the undesirable area law contribution; we return to such a construction in Sec.~\ref{subsec:MI}.

\subsection{Removing trivial boundary contributions} \label{subsec:MI}

\begin{figure}
    \centering
    \includegraphics[width=0.8\linewidth]{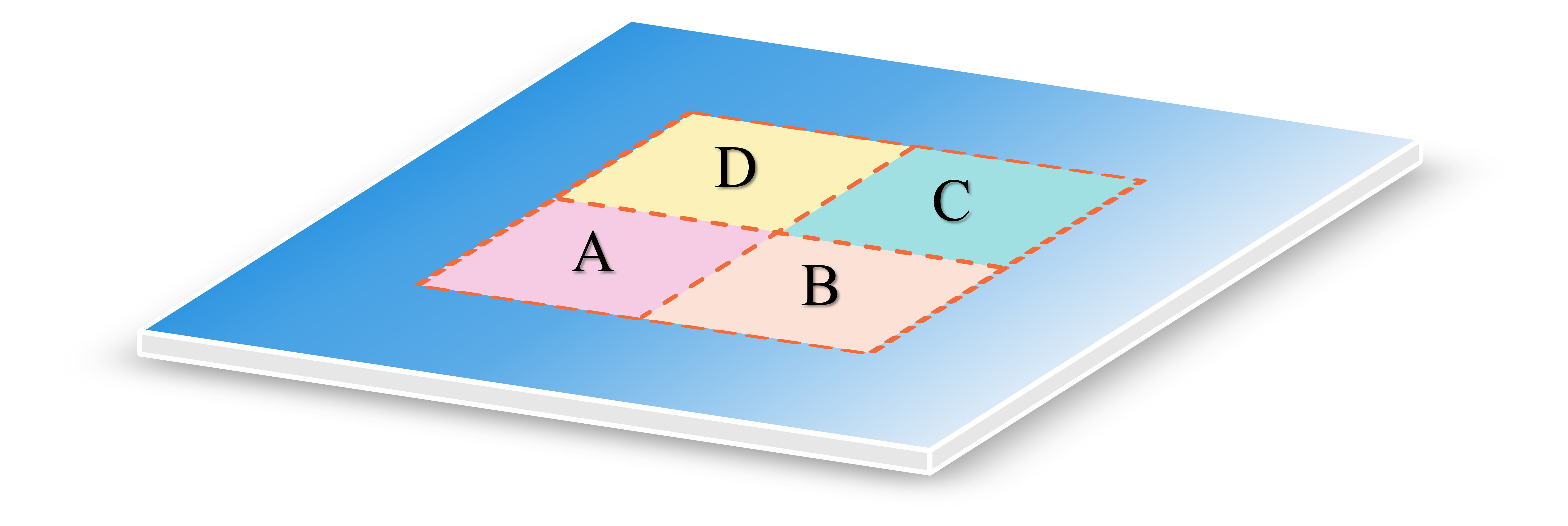}
    \caption{Schematic of the partition of the subsystem used for obtaining entanglement quantities with trivial boundary contributions removed. An initial bipartition into $\Omega$ and $\bar \Omega$ is defined, and $\Omega$ is then further split into a quadripartition as shown. 
     }
    \label{fig:picture2}
\end{figure}

\begin{figure}
    \centering
    \includegraphics[width=1\linewidth]{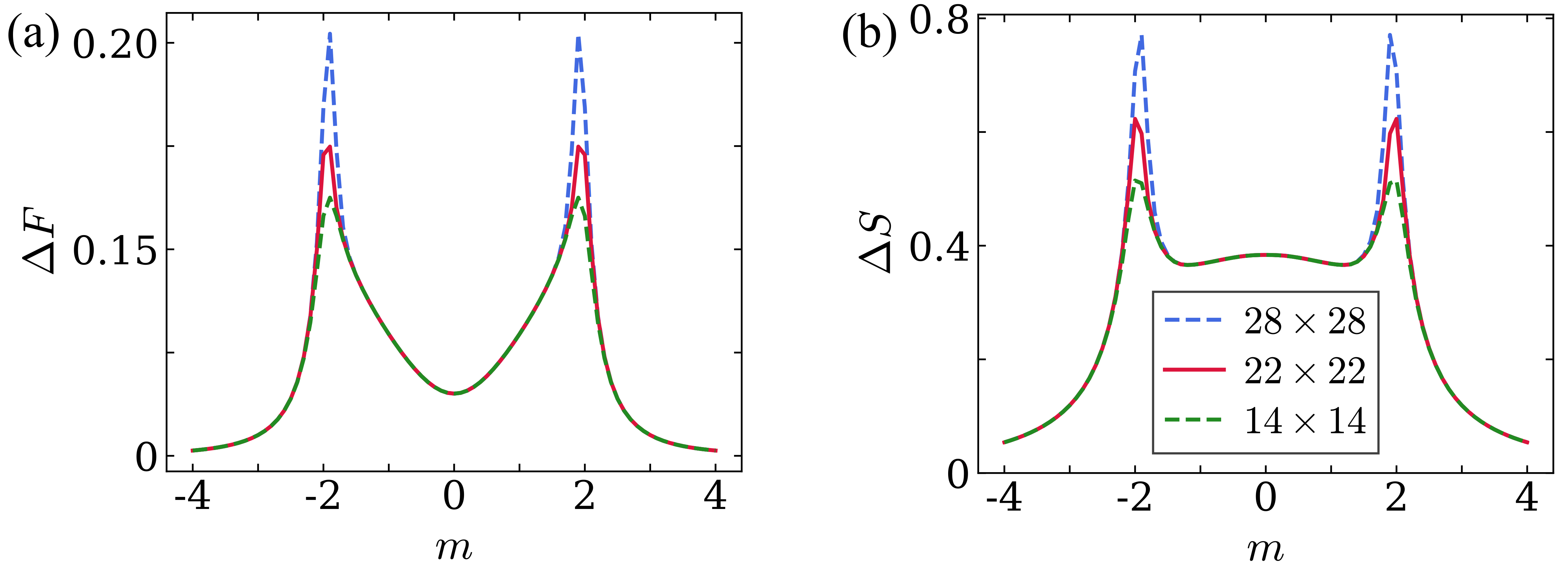}
    \caption{(a) Modified fluctuations $\Delta F = \mathrm{Var}N_{\text{AB}} + \mathrm{Var}N_{\text{AD}} - \mathrm{Var}N_{\text{AC}} + \mathrm{Var}N_{\text{C}}  -\mathrm{Var}N_{\text{ABD}}$ for the setup in Fig.~\ref{fig:picture2} as a function of $m$, with system size $L = 30$ and three different subsystem sizes $(L_{\Omega}=14, 22, 28)$ shown. The fluctuations are generally subsystem size invariant, and clearly take a higher value in the topological phase; at the phase transition, where the gap closes, size dependent peaks can be observed. (b) Modified entanglement entropy $\Delta S = S_{\text{AB}} + S_{\text{AD}} - S_{\text{AC}} + S_{\text{C}} - S_{\text{ABD}}$ for the same setup. Quantitative differences can be observed, but the qualitative features are similar.}
    \label{fig:quadpart_allsize}
\end{figure}

As seen in the previous section, the bipartite entanglement entropy in itself is not a very clear indicator of topology in this system, and as such, being used as a proxy for it, neither are the fluctuations. This is due to being obscured by entanglement across the boundaries of the system, which will be present even for topologically trivial systems, and presumably scale with the size of the system boundary. This problem has previously been addressed by the concept of topological entanglement entropy \cite{TEE_2006,Levin_2006}. By choosing combinations of entanglement entropies of subsystems with prefactors to eliminate boundary contributions, the topological contribution to entanglement can be isolated. One similar scheme adapted for second-order topological insulators, consisting of dividing the system into four subsystems each locally respecting the symmetries of the Hamiltonian, was previously proposed in Ref.~\cite{wang_2018}; by dividing the full system into four subsystems $A, B, C, D$, respecting the symmetries of the Hamiltonian, the topological contribution to the entropy can be obtained. However, we note that this approach would typically require measuring the full state of the system, which quickly becomes experimentally infeasible as system size increases. Hence, it would be preferable to instead apply a similar treatment to a smaller subsystem of the whole. This can be achieved by a simple modification of the scheme of Wang \textit{et al.}, wherein additional terms are considered in order to remove edge contributions from entanglement with the thus-introduced subsystem complement $\overline{ABCD}$. To this end, we first divide the system in a bipartite manner into $\Omega, \bar \Omega$ as previously, then further propose a quadripartite construction $\Omega = A\cup B\cup C\cup D$, similarly to Ref.~\cite{wang_2018}, as shown in Fig.~\ref{fig:picture2}; here, however, the quadripartition is only of the subsystem, not the full system. We then define a modified entropy scheme through $\Delta S = S_{AB} + S_{AD} - S_{AC} + S_{C} - S_{ABD}$ to remove contributions from topology across the boundaries of the various subsystems. As $\Omega$ generically does not cover the full lattice and may entangle with $\bar \Omega$ (blue area in Fig.~\ref{fig:picture2}), the terms $S_{C} - S_{ABD}$ are added to remove any contribution from entanglement across that initial bipartition. In the limit where $\Omega$ covers the entire system, we have $\bar C = A\cup B \cup D$ and these terms cancel out. The procedure here is illustrated in Fig.~\ref{fig:quadpart_allsize} for different subsystem $ABCD$ sizes. With the removal of the trivial boundary contribution,  the phase transition is readily observed as a peak in the topological entanglement entropy, as shown in Fig.~\ref{fig:quadpart_allsize}(b). The height of these peaks depend on the size of the subsystem, while the remainder of the curve should remain approximately constant given that the subsystem is large compared to the correlation length of the system. 

Experimentally, the topological entanglement entropy is certainly not easier to observe than the bipartite entanglement entropy, and as such is not a convenient indicator of the topology of a system. However, an identical scheme can be proposed for use with the particle number variance of each individual subsystem, with $\Delta F \equiv \mathrm{Var}N_{AB} + \mathrm{Var}N_{AD} - \mathrm{Var}N_{AC} + \mathrm{Var}N_{C}  -\mathrm{Var}N_{ABD}$. The theoretical underpinning of mutual fluctuations and their generalization is not quite as clear as for entropy, lacking some properties such as subadditivity \cite{moghaddam_2023}; however, to the extent fluctuations scale with the entanglement entropy, a distinction between trivial and topological phases can nevertheless be expected, as while the prefactor for each term may vary the overall scaling may be expected to follow phase boundaries as it does for the entropies. This is also observed here, as shown in Fig.~\ref{fig:quadpart_allsize}(a); the modified fluctuations essentially replicate the curve of the modified entropy, including, crucially, the clear peaks signifying the phase transition points. Otherwise the overall shape here is indeed not identical, for reasons mentioned previously; however, we see that each graph is essentially independent of subsystem size excepting the phase transition peaks. The shape is also to some extent tolerant of the presence of disorder or nonzero temperature, as discussed in Appendix~\ref{app_A}. Experimentally obtaining this curve should require no more than measuring the expected particle fluctuations of each subsystem combination and adding them appropriately.  As only the peaks at phase transition show any significant dependence on subsystem size, in principle the curve can be obtained by measuring a reasonably small subsystem regardless of the size of system, only needing to be large compared to the correlation length of the corner modes, as well as distant from any boundaries of the system. However, there is a size constraint in the opposite direction; the fluctuations of each subsystem will, as previously noted, scale linearly with the length scale of said subsystem; hence, obtaining $\Delta F$ from larger subsystems will require accurately subtracting ever larger contributions from each other. The optimal system size will thus be limited from below by the theoretical bound presented by the scale of corner modes, and from above by practical concerns related to the experimental accuracy of the fluctuation measurements. As the fluctuations only scale with the size of the subsystems, the rest of the system can be arbitrarily large, which has some benefits. For infinitely large systems, or with periodic boundary conditions, the four-square configuration is not necessary, as any or all of the terms in the sums defining $\Delta F$ (or $\Delta S$) can be replaced by a value measured in some arbitrary identical subsystem located elsewhere in the system. For smaller systems with open boundary conditions, however, deviating from the specified configuration will somewhat alter the shape of the $\Delta S$ and $\Delta F$ combinations as the subtraction of terms of similar magnitude enhances even small shifts.

\subsection{Dynamical entanglement properties}
Previous works have suggested that signatures of topology could also be observed in the time-dependent behavior of the entanglement as the system undergoes a quench crossing a phase boundary \cite{Heyl_2013, Carl_2016, Caio_2015, Vajna_2014, Flaschner_2018, Jurcevic_2017, bauer_2025}. In 1D topological systems, it was found that transitions where the post-quench time evolution is governed by a topological Hamiltonian correspondingly show signatures in the time evolution of entanglement quantities \cite{poyhonen_2021}. Quenches have also been studied for the specific case of second-order topological insulators in 2D: in particular, a previous work focusing on higher-order systems found that a quench enacting a transition from a trivial to a higher-order topological state will result in corresponding signatures in the Loschmidt echo of the system and its derivative \cite{maslowski_2023}. It is worth considering whether the time-dependent behavior of other quantities could also be used to detect such a transition.

Guided by these previous results, we will consider the time evolution of entanglement quantities in this model following a quench. Pioneered by Peschel \cite{peschel_2003, peschel_2007, peschel_2009}, to study non-equilibrium dynamics following a sudden parameter quench, it is appropriate to calculate the time-dependent correlation matrix $C_{lm}(t) = \langle c^\dagger_l(t) c_m(t) \rangle$, similarly to the static case earlier, using post-quench operators but taking the expectation value in terms of the pre-quench ground state. By expanding the post-quench operators in the basis of the pre-quench eigenstates, the time evolution of the subsystem can be constructed dynamically. 

\begin{figure}
    \centering
    \includegraphics[width=0.8\linewidth]{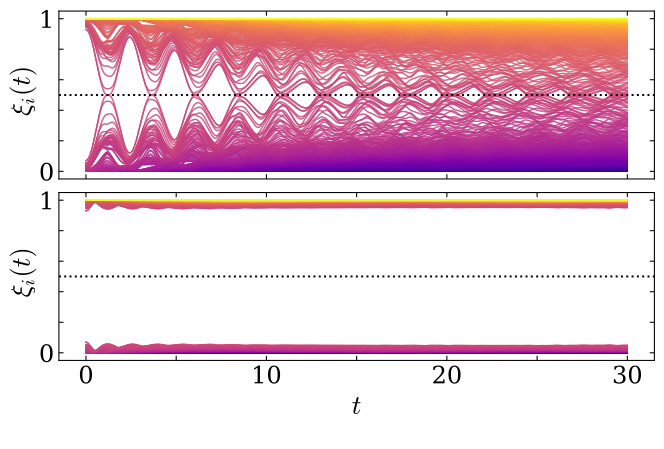}
    \caption{The time evolution of the spectrum of the correlation matrix for system size $L=30$ and subsystem size $L/2$. Top: A quench of trivial to topological parameter regime: $m_i=2.5$ and $m_f=0.5$. Bottom: A quench of trivial to trivial parameter regime: $m_i=2.5$ and $m_f=4$.}
    \label{fig:quenchspec}
\end{figure}

\begin{figure}
    \centering
    \includegraphics[width=1\linewidth]{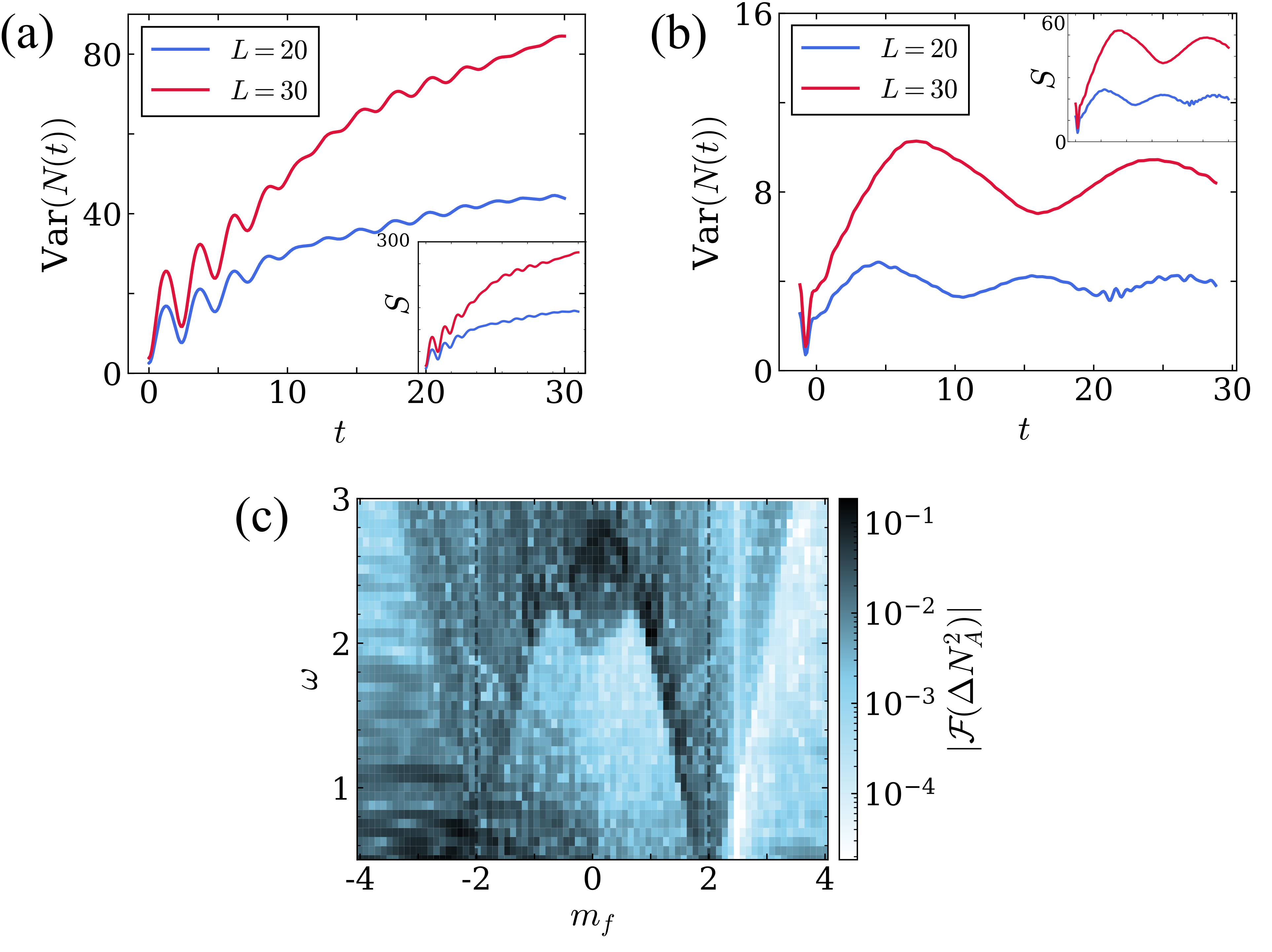}
        \caption{(a) Time-dependent bipartite particle number variance and entanglement entropy for a quench of trivial-to-topological parameter regime: $m_i=2.5$ and $m_f=0.5$, shown for two system sizes ($L=20$ and $L=30$) at fixed subsystem-to-system ratio ${L_{\Omega}}/{L}=1/2$; the oscillation frequency is unchanged between two sizes. (b) Same as (a), a quench of trivial-to-trivial parameter regime: $m_i=2.5$ and $m_f=4$, included as a control. 
        (c) Amplitude of the Fourier transform of the particle number variance, $|\mathcal{F}(\Delta N_A^2)|$, for a system size $L=30$ and subsystem size $L_{\Omega}=15$, computed over the late-time interval ($100\leq t \leq 200$) after subtracting the temporal mean and applying a Hann window, as a function of frequency $\omega$ and post-quench mass $m_f$, for a fixed pre-quench mass $m_i = 2.5$ in the trivial phase. The dashed lines indicate the phase boundaries $m_f = \pm 2$. A band of enhanced spectral weight ("arc") is concentrated within the topological window $|m_f|<2$, narrowing and decaying smoothly as $m_f$ approaches the phase boundaries, with frequencies independent of subsystem size. At low frequencies, a band of higher intensity, very weakly dependent on $m_F$, is also present, but is less informative with regards to topology, and the frequency axis has been restricted to $\omega \geq 0.5$ to exclude its contribution, which would otherwise obscure the more interesting features. 
        } 
        
    \label{fig:quench}
\end{figure}

In Fig.~\ref{fig:quenchspec}, we plot the time evolution of the correlation matrix spectrum for quenches of trivial-to-topological and trivial-to-trivial regimes. While the former (top) exhibits several gap closings, the latter (bottom) follows a monotonous pattern, with no closing.
The time-dependent behavior of the particle number fluctuations and bipartite entanglement entropy are shown in Fig.~\ref{fig:quench}, and display oscillations on timescales governed by the spectra in Fig.~\ref{fig:quenchspec}. In particular, the time evolution in Fig.~\ref{fig:quench}(a) looks notably similar to the results obtained for a topological (first-order) system in Ref.~\cite{poyhonen_2021}. The oscillatory behavior is present for a quench from the trivial-to-topological parameter regime, with the frequency independent of system size. However, the data in Fig.~{\ref{fig:quench}} (b), for a quench within the trivial regime as a control case, shows weaker and less structured oscillations by comparison, unlike the sharper suppression reported in the Supplemental Material of Ref.~\cite{poyhonen_2021} for a trivial quench in 1D. To more fully characterize the bipartite dynamical behavior, we compute the amplitude of the Fourier transform of the particle number variance over a late-time interval ($100\leq t \leq200$), as a function of angular frequency $\omega$ while sweeping the post-quench mass $m_f$ at fixed pre-quench mass $m_i = 2.5$, as shown in Fig.~\ref{fig:quench}(c). Before transforming, we subtract the temporal mean of this interval and apply a Hann window to suppress spectral leakage. Restricting the Fourier transform to the window of time evolution as described suppresses the contribution of the initial transient, during which the particle number fluctuation evolves on a non-stationary background following the quench. The general features here do not strongly depend on the time window chosen once the first peak has been omitted, as long as the system is clean; however, in the presence of disorder, oscillations at later times may be suppressed, as noted in Appendix~\ref{app_A}. A distinct region of enhanced spectral intensity, an "arc", is concentrated within the topological phase window ($|m_f|<2$), narrowing and decaying smoothly as $m_f$ the approaches and crosses the phase boundaries. We also find the low-frequency weight localized around $\omega \approx 0$, which appears across all $m_f$, and is as such less useful in specifying a topological effects; it may reflect a number of slower changes, including things like system size and geometry-dependent variations as well as non-oscillatory relaxation of the particle number variance, of which we will not attempt a detailed study in this work. As such, we omit the low-frequency data from the figure, focusing on the arc which distinguishes a quench crossing phase boundaries. However, while this pattern appears to hold for this particular model, we find that, as a signature of higher-order topology in two dimensions, these oscillations are not a reliable measure. It is possible to construct topologically trivial models that feature similar oscillations to the trivial-to-topological case here;
in Appendix~\ref{app_B}, we show an example of the quench dynamics of a non-topological Hamiltonian, where, although the time behavior of the entanglement spectrum displays no crossings in the entanglement spectrum, the time-dependent particle number variance and entanglement entropy nevertheless feature oscillations similar to those observed here. As such, data from the quench in isolation is insufficient to determine topological character; however, the oscillations could nevertheless be used as an additional confirmation once other approaches, such as the fluctuation constructions outlined previously, indicate nontrivial topology is present.

\section{Conclusions and discussion}
We examined the entanglement properties of a higher-order topological insulator and discussed some potentially experimentally viable approaches to using entanglement effects as a probe of the system topology. Our results indicate that the particle number fluctuations correspond very well to the von Neumann entanglement entropy in this system, providing a method of accessing some features of the entanglement without requiring a full state tomography or more complicated measurement protocols. The bipartite quantities, while exhibiting some dependence on topology, were found to be insufficient as an indicator of topological phase in contrast to 1D topological systems. However, we found that a suitable combination of entanglement entropies obtained by further dividing a bipartite subsystem into a quadripartite construction can be used to probe the system topology, and further that the same setup using particle number fluctuations rather than entropy retains the crucial features needed to do this. The main idea of measuring the particle number fluctuations of subsystems is, we believe, feasible with current experimental techniques \cite{Song_2012,kaufman_2016}. While there may not be an extant material featuring the exact Hamiltonian described here, experimental realizations have also been realized on different platforms, such as photonic systems or in ultracold atom gases \cite{Yan_2022, pelegri_2019}. However, it is important to note that we have here only studied one type of transition, from trivial to second-order insulator. While that is perhaps the most relevant case -- showing that nontrivial topology is present -- transitions to different topological phases would likely be harder to detect using the methods outlined here, and further study is needed to see whether this approach could be suitable in such cases. Adapting the idea to higher-order topological insulators with different crystalline symmetries, even those with the same type of transition, could also require some study in order to obtain a suitable combination of subsystems.

\section{Acknowledgements} 
The authors acknowledge the Finnish Research Council project 363879.

\section{Author contributions}
The numerical simulations were carried out by M.M.F. The project was planned, the results were analyzed and the manuscript was prepared jointly by both authors. \\

\appendix

\section{Robustness of entanglement} \label{app_A}
The results in the main text were obtained strictly for clean systems at absolute zero. As a nonzero temperature affects fluctuations, and disorder will necessarily also be present in experiments -- often breaking the $C_4T$ symmetry protecting the topology -- it is necessary to consider whether the results here can be obtained under less ideal circumstances. 

\begin{figure}
    \centering
    \includegraphics[width=1\linewidth]{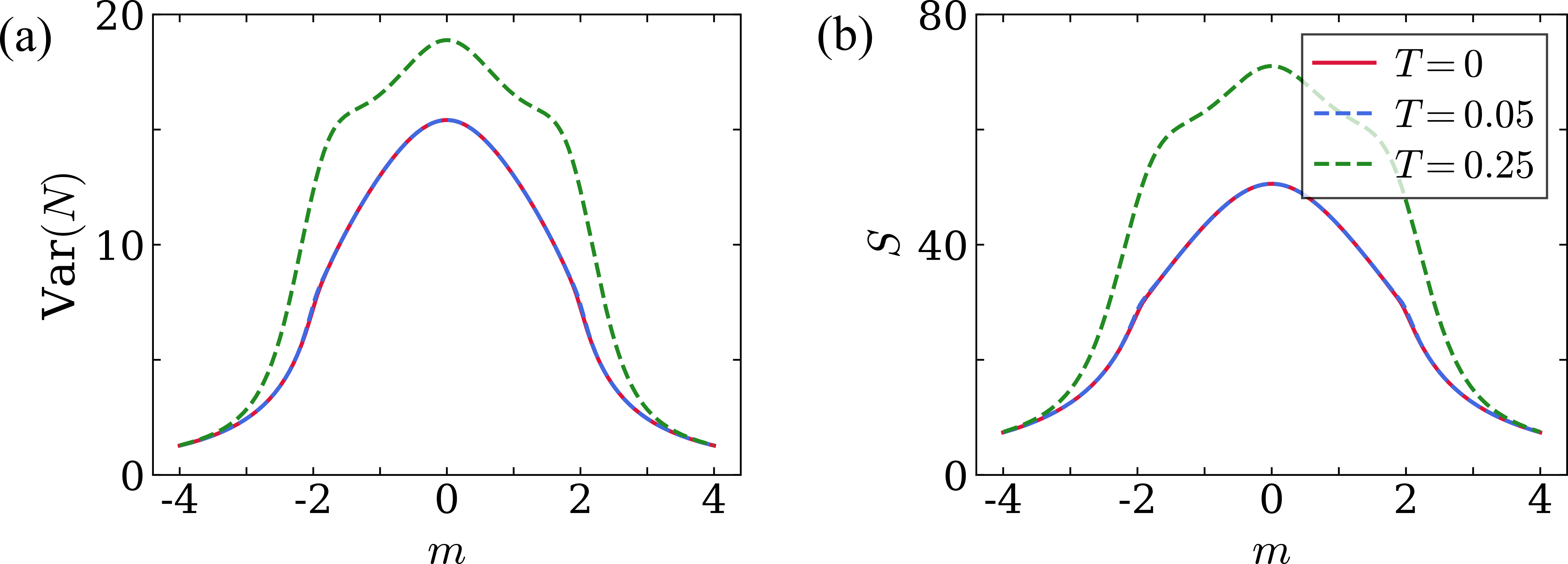}
    \caption{Bipartite (a) fluctuations and (b) von Neumann entanglement entropy as a function of $m$ for selected temperatures, for a system size $L=30$, subsystem size $L_\Omega=15$, and model parameters $t=0.5, \lambda=0.5, \Delta=0.25$. The plot assumes natural units with $k_B = 1$, and so $T$ (equivalently $\beta = 1/T$) is measured in terms of the Hamiltonian parameters. }
    \label{fig:bipart_T}
\end{figure}

\begin{figure}
    \centering
    \includegraphics[width=1\linewidth]{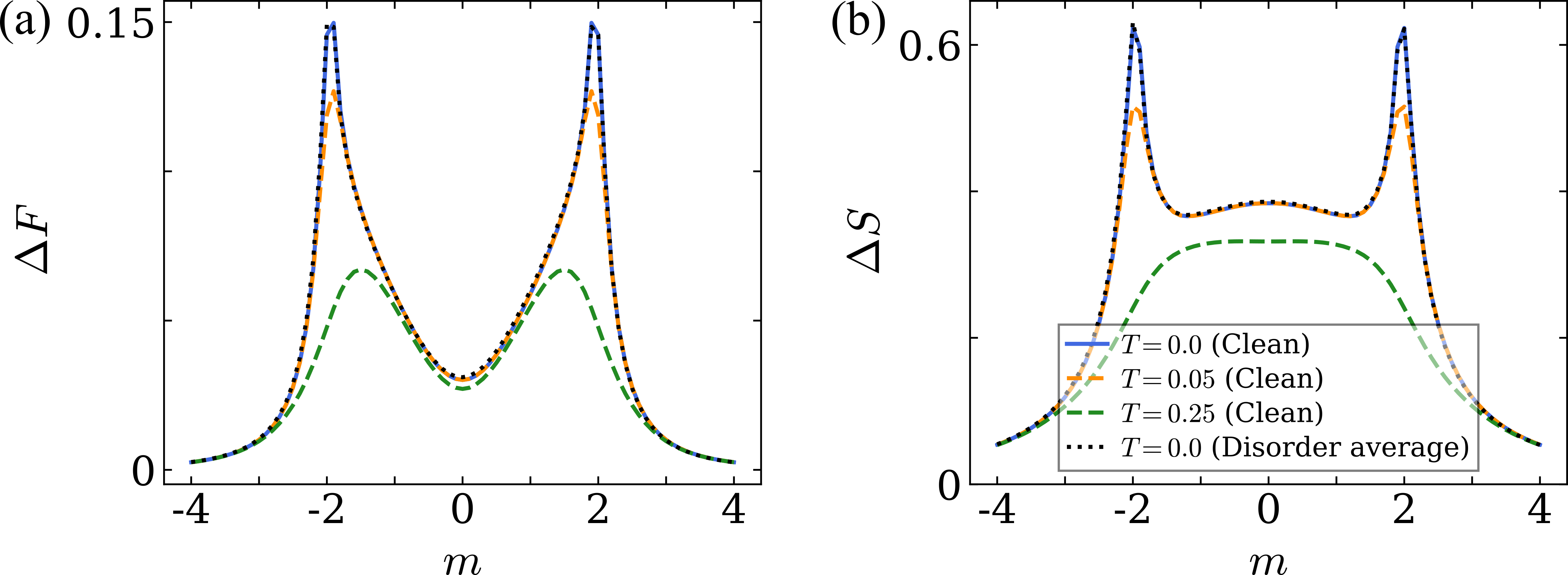}
    \caption{The effect on (a) $\Delta F$ and (b) $\Delta S$ as nonzero temperature or $C_4T$-breaking onsite disorder are introduced, for a system size $L=30$ with the quadripartite subsystem size $L_{\Omega}=22$ and model parameters $t=0.5, \lambda=0.5, \Delta=0.25$. Solid and dashed curves show the clean $(W=0)$ results at $T=0.0, 0.05, 0.25$. The dotted curve shows the $T=0$ result averaged over $10$ independent disorder realizations $(W=0.2)$, each held fixed across the full sweep of $m$. The disorder-averaged plot is mostly indistinguishable from the clean result. }
    \label{fig:quadpart_disT}
\end{figure}

In Fig.~\ref{fig:bipart_T} we show the bipartite fluctuations and entanglement entropy as a function of $m$ for three different temperatures.  We work in units where $k_B = 1$, such that $T$ (equivalently $\beta = 1/T$) is measured in the same energy units as the Hamiltonian parameters. A low temperature of $T = 0.05$ essentially does not affect the plot, while at $T = 0.25$ the curve is shifted to somewhat higher values and its shape displays new features. In this particular case the phase boundary may even be visually enhanced at moderate temperatures, but we do not necessarily claim this is a generic feature. To further test the robustness of the signatures, we introduce static, spatially uncorrelated disorder into the mass term, $m(\vec{r})=m+\delta m(\vec{r})$, with $\delta m(\vec{r})$ drawn from the uniform distribution over the interval $(-W, W)$. In Fig.~\ref{fig:quadpart_disT}, we show $\Delta F$ and $\Delta S$ for nonzero temperatures or $C_4T$-breaking onsite disorder. Solid and dashed curves show the clean $(W=0)$ result at $T=0.0, 0.05, 0.25$. The dotted curve shows the $T=0$ result averaged over $10$ independent disorder realizations $(W=0.2)$, each held fixed across the full sweep of $m$. At this disorder strength, the disorder-averaged plot is mostly indistinguishable from the clean result, which confirms the robustness of the subsystem quadripartite construction against onsite disorder. We note that the disorder here does break the rotational symmetry of the system, and as such removes the topological protection, but that is likely to also be the case for any experimental disorder; nevertheless weak disorder, as seen, does not obscure the signal. By contrast, increasing temperature broadens and suppresses the peaks, most notably at $T=0.25$. Generally, as temperature increases its contribution to fluctuations can be expected to dominate the topological features, also providing different subsystem scaling; for sufficiently low temperatures and small subsystems, as seen here, the topological contribution is more prominent. 

\begin{figure*}
    \centering
    \includegraphics[width=0.8\linewidth]{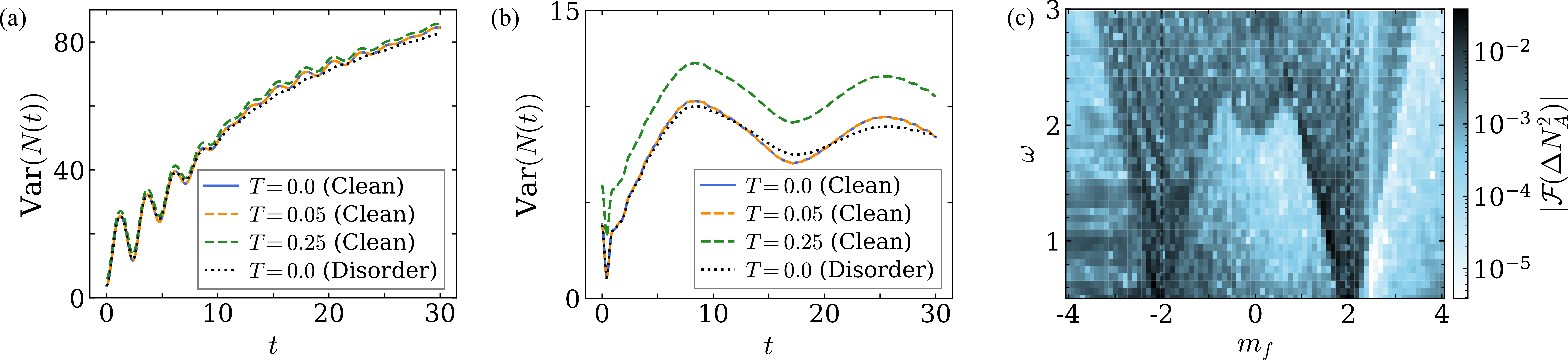}
    \caption{Time-dependent bipartite (a) a quench of trivial-to-topological parameter regime: $m_i=2.5$ and $m_f=0.5$ and (b) a quench of trivial-to-topological parameter regime: $m_i=2.5$ and $m_f=4.0$, for a system size $L=30$, subsystem size $L_{\Omega}=15$, and model parameters $t=0.5, \lambda=0.5, \Delta=0.25$, at selected temperatures in clean case $(W=0)$ and at $T=0$ in the disordered case $(W=0.2)$. (c) Amplitude of the Fourier transform of the particle number variance of the disordered case, $|\mathcal{F}(\Delta N_A^2)|$, as a function of frequency $\omega$ and post-quench mass $m_f$, for a fixed pre-quench mass $m_i = 2.5$ in the trivial phase.}
    \label{fig:disorder}
\end{figure*}

Furthermore, in Fig.~\ref{fig:disorder} we illustrate the effect of nonzero temperatures or disorder on time evolution entanglement quantities following a quench. Here, the same single disorder realization is held fixed in both the pre- and post-quench Hamiltonians, so that the quench remains a sudden change in the uniform mass only. We compare $W = 0.2$ against the clean $(W=0)$ case at otherwise identical parameters, using the same time evolution and entanglement quantities described above; as before, $T=0.05$ is indistinguishable from the clean result; while $T=0.25$ produces a modest but visible shift in both quantities throughout the evolution. As shown in Fig.~\ref{fig:disorder}(a,b), this level of disorder leaves the qualitative dynamical signatures, particle number variance and entanglement entropy, almost unchanged for both trivial-to-topological and trivial-to-trivial quenches at early times, despite the fact that this disorder breaks the symmetries protecting the higher-order phase. However, at longer times the higher-frequency oscillations in the trivial-to-topological quench are dampened by the disorder. Stronger disorder may of course result in much less predictable signals even at early times, but is also otherwise undesirable given the symmetry breaking. Temperature has a less destructive impact on the time-dependent signal, mainly acting as a constant shift towards higher fluctuations, presumably because the quench has been assumed to not affect the overall temperature of the system. Finally, Fig.~\ref{fig:disorder}(c) shows the amplitude of the Fourier transform of the late-time particle number variance of the disordered case $(W = 0.2)$, $|\mathcal{F}(\Delta N_A^2)|$, as a function of frequency $\omega$ and post-quench mass $m_f$, for a fixed pre-quench mass $m_i = 2.5$ in the trivial phase. As expected, comparing Fig.~\ref{fig:disorder}(c) to the clean result (Fig.~\ref{fig:quench}(c)), disorder introduces more broadband noise across the spectrum, degrading the contrast and sharpness of the arc; while a similar envelope near $|m_f|\lesssim 2$ remains detectable, the clean and well-defined arc structure seen in the absence of disorder is somewhat obscured at this disorder strength. 


\section{Contrast with a topologically trivial Hamiltonian} \label{app_B}
\begin{figure}
    \centering
    \includegraphics[width=1\linewidth]{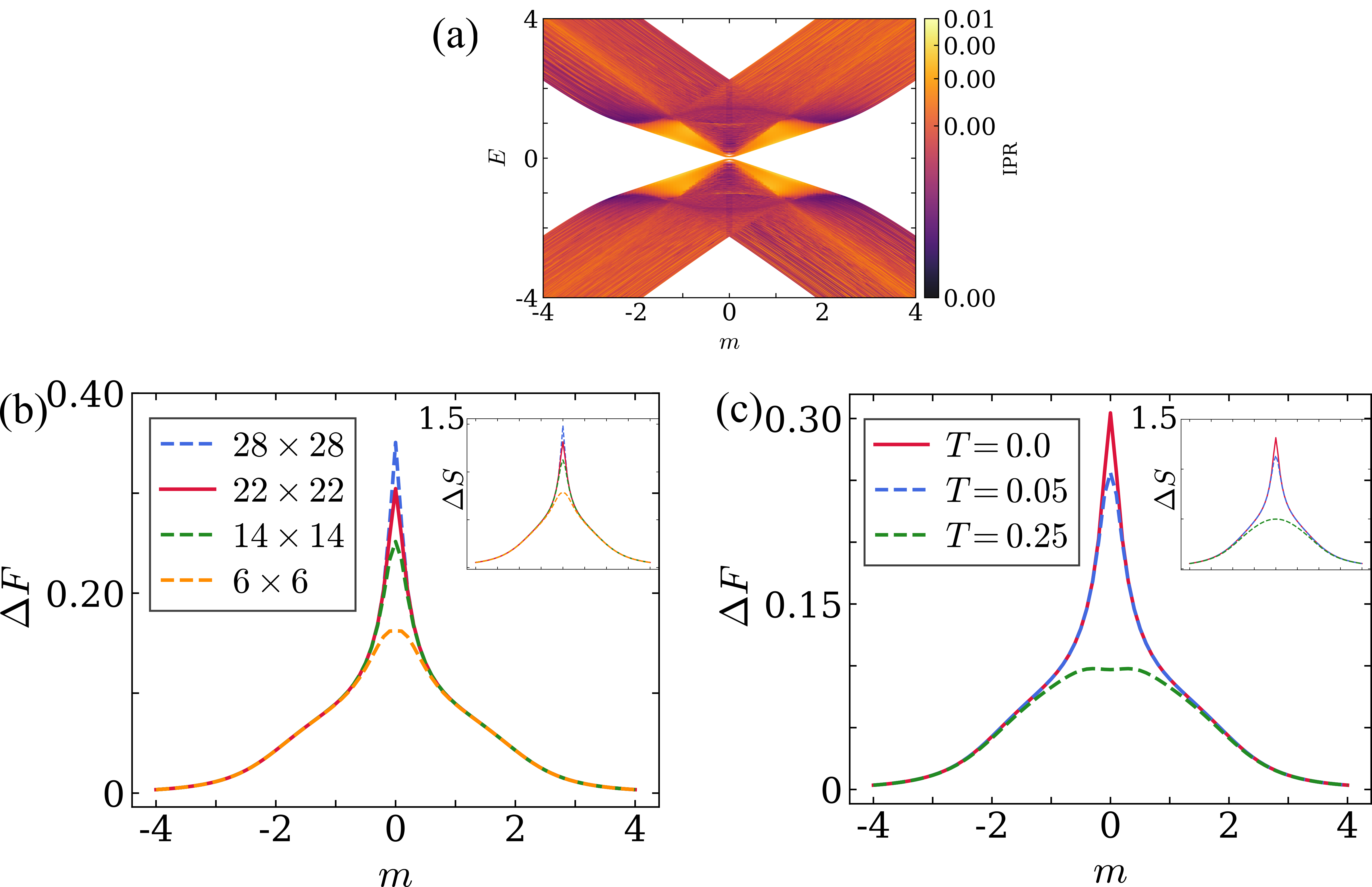}
    \caption{(a) Energy spectrum of the gapped, non-topological Hamiltonian as a function of the parameter $m$ for a system size $L = 30$. The eigenvalues have been colored by the inverse participation ratio $\text{IPR} = \sum_{\vec r_j} |\psi({\vec r}_j)|^4$. (b) $\Delta F$ and $\Delta S$ for the gapped, non-topological Hamiltonian as a function of $m$ for four different subsystem sizes $L_{\Omega}$. (c) $\Delta F$ and $\Delta S$ for the gapped, non-topological Hamiltonian as a function of $m$ at $T = 0.0, 0.05, 0.25$ for system size $L =30$ and subsystem size $L_{\Omega}=22$.} 
    \label{fig:static_gap}
\end{figure}

It is straightforward to produce a topologically trivial Hamiltonian with comparable parameters to the model of the main text. This is achieved by taking the sign of the $\Delta$ term in Eq.~\eqref{eq:hamiltonian_c4} to be independent of direction, $\Delta \cos(2\theta_a) \to \Delta$ for $a = x, y$, rather than alternating between $x$ and $y$. This leaves the bulk gapped throughout the parameter range considered, excepting a gap closing at $m = 0$, as seen in Fig.~\ref{fig:static_gap}(a); the model is topologically trivial for all $m$, with no phase transition as $m$ is varied, but does feature some states located around the edges at nonzero energy when the mass is small. While certainly not representative of all trivial systems, it provides a useful point of comparison to the topological case of the main text.

\begin{figure}
    \centering
    \includegraphics[width=1\linewidth]{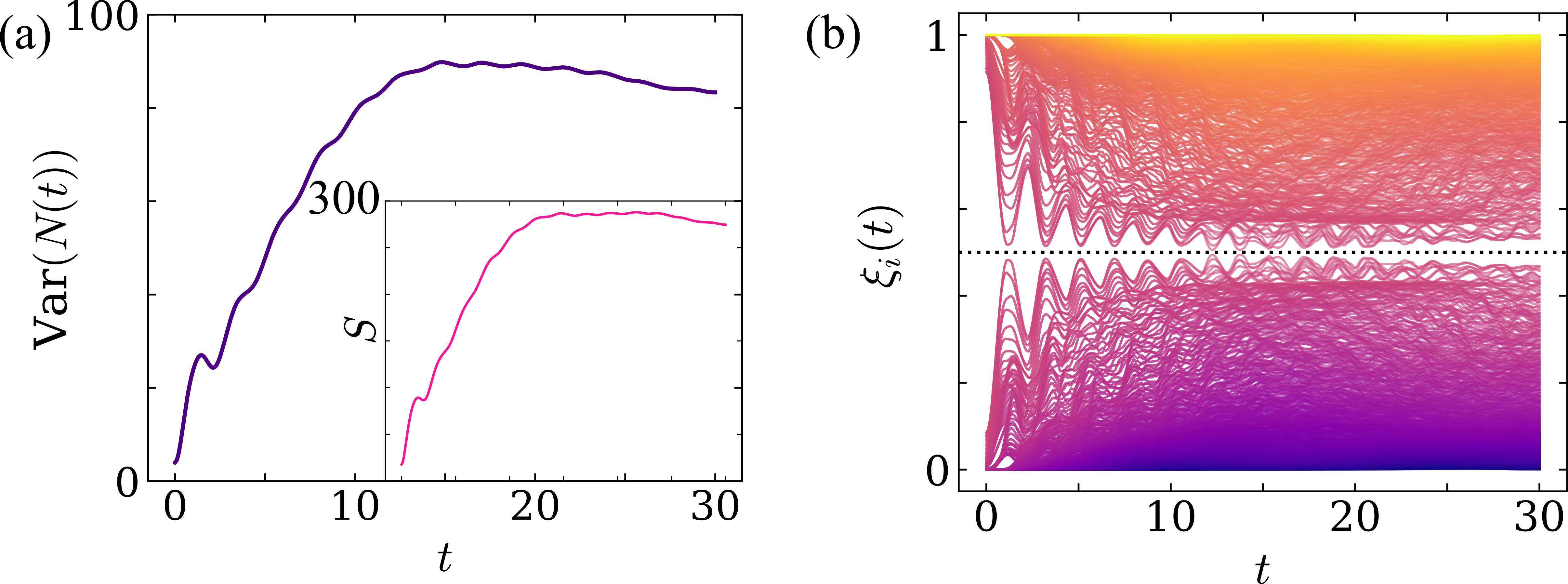}
    \caption{Time-dependent bipartite (a) particle number variance and entanglement entropy of the gapped, non-topological Hamiltonian, for a system size $L=30$, subsystem size $L_{\Omega}=15$, and model parameters $t=0.5, \lambda=0.5, \Delta=0.25$ and (b) entanglement spectrum for a quench of trivial-to-topological parameter regime: $m_i = 2.5$ and $m_f=0.5$}
    \label{fig:quench_gap}
\end{figure}

Fig.~\ref{fig:static_gap}(b) and (c) show $\Delta F$ and $\Delta S$ as a function of $m$ for a fixed system size with four different subsystem sizes $L_\Omega$ and a fixed size scale at three different temperatures, respectively. In both cases, no sign of a topological phase is present, as expected, though the results are peaked around $m = 0$ where the gap closes, with the peak scaling with subsystem size. However, we can also examine the bipartite quench dynamics of this Hamiltonian, using the same mass values, $m_i = 2.5$ and $m_f=0.5$, as the quench from the trivial-to-topological parameter regime discussed in the main text. As shown in Fig.~\ref{fig:quench_gap}, the time-dependent particle number variance and entropy loosely resemble the results of the topological Hamiltonian quench in both scale and shape,
which illustrates their limited value as standalone diagnostics. The curves, despite not being identical to the model of the main text, do showcase essentially similar oscillations, at a comparable frequency and independent of subsystem size, and in an experimental setting the quantitative differences would certainly be muddled further. While this particular model was in principle constructed from a topological insulator, and has somewhat edge-localized states as a result, the system is nevertheless topologically trivial. Hence, the quench fluctuations are not reliable evidence of a topological phase in the system, though they may still serve as an additional test when topology has first been found through a different method. We do note that despite the otherwise similar quench, the gap in the spectrum of the correlation matrix, as seen in Fig.~\ref{fig:quench_gap}, never quite closes, as is expected given the lack of a phase transition here.

\section{Periodic Boundary Condition} \label{app_C}
In the main text, all the results are obtained for a system with open boundary conditions and with a subsystem that is not located at the system edge. This was chosen in order to be more directly applicable to experiments, in which periodic boundary conditions (PBC) may be difficult to achieve. In the main text we also claim that, as must necessarily be true for a useful measure, the results largely do not depend on the boundary conditions of the system, as long as the edge is far away from the subsystem. 

\begin{figure}
    \centering
    \includegraphics[width=1\linewidth]{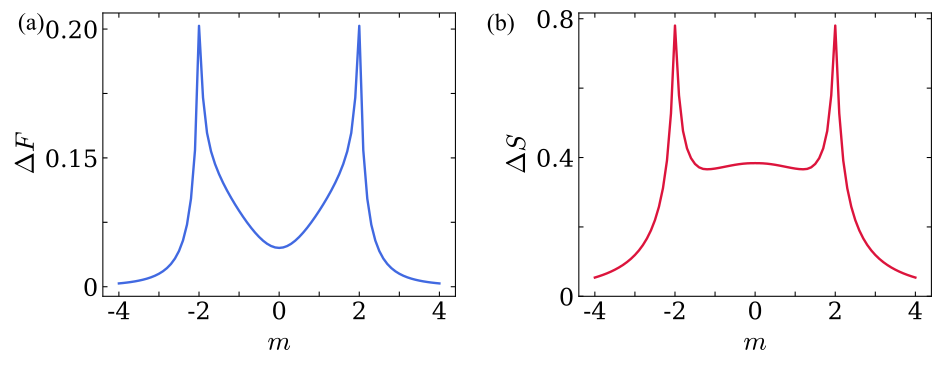}
    \caption{(a) $\Delta F$ and (b) $\Delta S$ as a function of $m$ for the PBC Hamiltonian, for a system size $L=30$, subsystem size $L_{\Omega}=22$. The parameters are the same as in Fig.~\ref{fig:quadpart_allsize} of the main text.}
    \label{fig:staticpbc2}
\end{figure}

\begin{figure}
    \centering
    \includegraphics[width=1\linewidth]{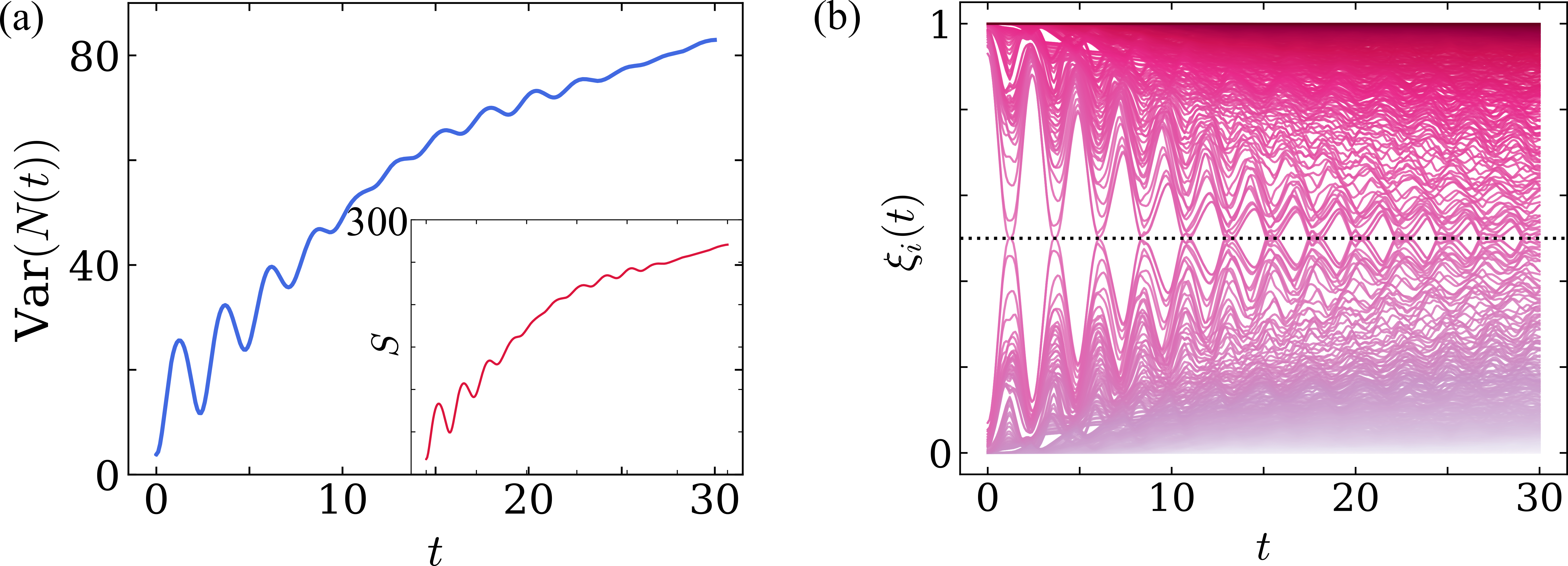}
    \caption{Bipartite time evolution of (a) fluctuation and entanglement entropy, for a system size $L=30$, subsystem size $L_{\Omega}=15$ and (b) entanglement spectrum for a quench of trivial-to-topological parameter regime: $m_i = 2.5$ and $m_f=0.5$ of the PBC Hamiltonian. The parameters are the same as in Figs.~\ref{fig:quenchspec} and \ref{fig:quench} of the main text.}
    \label{fig:quenchpbc}
\end{figure}

Some results for a system with periodic boundary conditions are shown in this Appendix. We find that the quadripartite subsystem modified fluctuations and entanglement entropy (Fig.~\ref{fig:staticpbc2}), and the time evolution of the entanglement entropy, particle number variance, and entanglement spectrum after a quench (Fig.~\ref{fig:quenchpbc}) are all in close quantitative agreement with the corresponding OBC results reported in the main text. This confirms that our results in the static case only depend on the local geometry of the subsystem and not on the boundary condition of the full system. For longer timescales the quench curves will indeed show a difference between PBC and OBC, but this is expected; as long as the timescale used is short compared to the time it takes for correlations to propagate from the system edge, this does not pose an issue.

\newpage

\bibliography{refs.bib}
\end{document}